# DiaRet: A browser-based application for the grading of Diabetic Retinopathy with Integrated Gradients

Shaswat Patel
*Department of Computer Science & Engineering*
*Netaji Subhas Institute of Technology*
New Delhi, India
shaswat178@gmail.com

Maithili Lohakare
*Department of Computer Science & Engineering*
*Pandit Deendayal Energy University*
Gandhinagar, India
maithililohakare@gmail.com

Samyak Prajapati
*Department of Computer Science & Engineering*
*National Institute of Technology Delhi*
New Delhi, India
pra.samyak@gmail.com

Shaanya Singh
*Department of Computer Science & Engineering*
*Pandit Deendayal Energy University*
Gandhinagar, India
shaan161200@gmail.com

Nancy Patel
*Bachelor of Medicine, Bachelor of Surgery*
*AMC MET Medical College*
Ahmedabad, India
patelnancynn@gmail.com

**Abstract—** Patients with long-standing diabetes often fall prey to Diabetic Retinopathy (DR) resulting in changes in the retina of the human eye, which may lead to loss of vision in extreme cases. The aim of this study is two-fold: (a) create deep learning models that were trained to grade degraded retinal fundus images and (b) to create a browser-based application that will aid in diagnostic procedures by highlighting the key features of the fundus image. In this research work, we have emulated the images plagued by distortions by degrading the images based on multiple different combinations of Light Transmission Disturbance, Image Blurring and insertion of Retinal Artifacts. InceptionV3, ResNet-50 and InceptionResNetV2 were trained and used to classify retinal fundus images based on their severity level and then further used in the creation of a browser-based application, which implements the Integration Gradient (IG) Attribution Mask on the input image and demonstrates the predictions made by the model and the probability associated with each class.

**Keywords: Diabetic Retinopathy, Deep Learning, Integrated Gradients, Computer-aided Diagnosis, Attention Mechanism, Explainable AI**

## I. INTRODUCTION

Diabetic Retinopathy is a disorder involving retinal capillaries of the retina of the eye seen in patients with diabetes mellitus (DM) caused primarily due to chronic hyperglycaemia (elevated blood sugar levels). Diabetic Retinopathy (DR) has resulted in securing the title of leading cause of ocular impairment and acquired blindness among adults in today's world. A total of 2.6 million people were affected with ocular impairment due to causes related to DR, a figure that is expected to rise to 3.2 million people in 2020, as reported by a study in 2015 [1]. Type 2 DM affected patients pose a lower risk of development of DR as compared to Type 1 DM patients. Time period of diabetes and poor metabolic control are major factors that are pivotal in the development of DR. Pregnancy, hypertension, smoking, anaemia, obesity and nephropathy are some conditions that can lead to the progression of DR.

DR diagnosis is done by examining retinal fundus images of affected patients, which show existence of certain types of lesions [2,3]. They can be categorized as follows:

- Microaneurysms (MA): These lesions are visible in the initial stages of DR and can be observed by clinical examination. They occur due to loss of pericytes (contractile cells) which leads to alterations of vascular intercellular contacts resulting in grape-shaped dilation of retinal capillaries known as microaneurysms.

- Haemorrhages (HM): These are lesions which appear as relatively larger spots on the retina when compared to other forms of lesions.

- Hard exudates (Hard EX): Leakage of lipids from abnormal blood vessels result in blurred vision in the central retinal area and appear as yellow (lipoproteinaceous deposits) spots on the retina.

- Soft exudates (Soft EX): These occur due to swelling present in the nerve fiber and are visible as white spots on the retina. They are also known as cotton wool spots due to their appearance.

DR can be mainly divided in two types, Non-Proliferative DR (NPDR) and Proliferative DR (PDR). NPDR can be observed by the emergence of microaneurysms, haemorrhages, soft exudates (cotton wool spots) and hard exudates as lipoproteinaceous deposits (yellow in colour). NPDR in later stages include an increased level of macular edema and vascular occlusion which is a restriction of blood supply to the retina. Neovascularization is considered to be the hallmark characteristic of PDR. The process of the rupture of microaneurysms leads to the activation of Vascular Endothelial Growth Factor (VEGF) which then leads to neovascularization (appears as tuft of capillaries), which then rupture again to activate VEGF, hence entering a vicious cycle. It can be spotted either at the disk (NVD) or elsewhere (NVE).

DR can be graded into 5 levels according to the severity present as illustrated in Figure 1, namely,

- No DR
- Mild DR
- Moderate DR
- Severe DR
- Proliferative DR

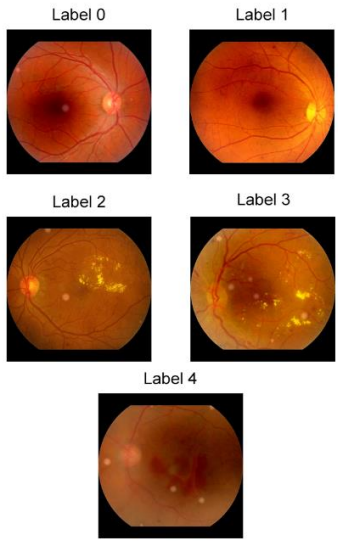

Fig. 1. The five DR stages: (a) unaffected normal retina (Label 0) (b) Mild DR (Label 1), (c) Moderate DR (Label 2), (d) Severe DR (Label 3), (e) Proliferative DR (Label 4) [3]

The rest of the paper is organized as follows: Section 2 consists of Background, Section 3 consists of Related Work, Section 4 explains about the Data Source, Section 5 consists of performance measures used, followed by the Proposed method in Section 6 and Section 7 consists of Results and Discussion.

## II. BACKGROUND

Early-stage detection of DR can be helpful for the prevention and reduction of the risk of permanent vision loss. Presently, the detection of diabetic retinopathy is done manually by a trained expert in the field of medicine by viewing retinal fundus images of affected patients. The manual DR detection system is therefore expensive in terms of time and cost as illustrated in Figure 2.

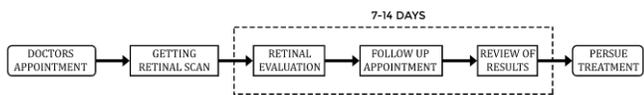

Fig. 2. Manual DR diagnosis pipeline [4]

Hence, the use of deep learning and computer-vision methods to assist physicians in a successful diagnosis is certain to be helpful. The solutions previously suggested for the detection of DR consists of feature extraction on retinal fundus images using standard machine learning algorithms which were trained on organised and curated retinal image datasets. However, detecting various features such as microaneurysms, haemorrhages, venous changes, cotton wool spots and hard exudates in the fundus images is difficult as compared to the feature extraction tasks used in traditional object-detection methods. The use of deep learning algorithms in recent times such as CNNs have been proven to be successful in computer-aided diagnosis, giving greater accuracy and achieving state-of-the-art results. The most important attribute of CNNs is that it helps in accurate feature extraction and detection which leads to correct prediction of severity grade levels.

The CNN architecture consists of three pivotal layers, Convolution layers (CONV), Pooling layers, and Fully Connected layers (FC). Convolution layers creates a convolution kernel that is merged with input layer to give outputs in the form of tensors. Pooling layers provide a way to down sample the features learned by the Convolution layers. The feature where a whole input image is described is taken care by the FC layers.

In this research work, we have used pretrained CNN architectures such as InceptionV3, InceptionResNetV2 and ResNet-50 that are trained on ImageNet dataset by implementing transfer learning to detect the severity levels of the fundus images.

In general, the strategy used in this research work consists of four basic steps, starting with i) collection of datasets, ii) preprocessing and degrading the images, iii) feeding the preprocessed and degraded images to the CNN models and iv) classification of the images. We trained and tested the CNN models on multiple datasets and further used the trained models to develop a browser-based application which takes a retinal fundus image as input and outputs the predicted severity grade of the image. This application involves the use of Integrated Gradients, which is used for showing the attribution of the input features to the output. This application is cost-effective, works well on real world data and would immensely aid medical professionals for successful diagnosis of DR.

## III. RELATED WORK

This section reviews previous substantial works done on this subject where the datasets were used to classify the images into categories related to severity level of the DR.

The Kaggle Dataset [5] was used by H. Pratt et al. [6] in the creation of a custom CNN model to classify the images into 5 DR severity levels. The preprocessing steps included image resizing to 512 x 512 pixels and colour normalization. Regularization (L2) and dropout methods were used to reduce overfitting, where 95% specificity, 75% accuracy and 30% sensitivity was obtained.

X. Wang et al. [7] used pretrained models of CNN, namely, VGG16 [8], AlexNet [9] and InceptionNet V3[10] on the Kaggle Dataset and were able to achieve 50.03% accuracy on average in VGG16, 63.23% in InceptionNet V3 and 37.43% in AlexNet. However, they attribute this to their use of just 166 images in training their models.

The works of Prasanna et al. [11] made use of a smartphone attached to a hand-held ophthalmoscope, which was used to directly capture the retinal fundus imagery and detect the symptoms of DR. They used four sets of images from the MESSIDOR Database [12] to develop a pattern recognition classifier. They were able to achieve an average sensitivity of 86% with the AUC of 0.844.

## IV. DATA SOURCE

We have made use of three publicly available datasets, namely, Indian Diabetic Retinopathy Image Dataset (IDRiD) [13], DDR dataset [14] and eyePACS (Kaggle) dataset for obtaining two-dimensional colour images of the retinal fundus. The images were obtained through extensive use of ophthalmoscopic imagery techniques conducted by retinal specialists. The datasets were utilized to train and test the deep learning models mentioned in the paper.

- Indian Diabetic Retinopathy Image Dataset (IDRiD): It contains total 516 images out of which it consists of 81 colour fundus images with signs of DR. Images for the left and right eye for each subject are provided.

- DDR dataset: It contains a total of 13,673 out of which it consists of 757 fundus images with signs of DR.

- EYEPACS (Kaggle) dataset: There are 88,702 images with various resolutions collected from domain experts. Only the ground truths of the training images are publicly available. Kaggle contains poor quality images and images with incorrect labelling.

The datasets consist of extremely noisy data and implementation of several pre-processing steps is required to get the images in usable format in order to train the models. The main aim is to come up with an effective algorithm that can perform well in the presence of noise and variations in the images.

## V. PERFORMANCE MEASURES

Performance measures are measures/metrics that are calculated for ML/DL methods or models in order to evaluate the performance of classifiers. The most commonly used performance measures are i) Accuracy, ii) Sensitivity, iii) Specificity and iv) Area under ROC curve (AUC).

Accuracy is the percentage of number of images the model predicted correctly, sensitivity tells us how precisely the model predicted the images, specificity tells us how many negative predictions were actually negative and AUC gives us measure of separability of distinguishing between different DR severity levels. Mathematical formulae of the metrics are:

$$Sensitivity = \frac{True\ Positives}{True\ Positives + False\ Negatives} \quad (1)$$

$$Specificity = \frac{True\ Negatives}{True\ Negatives + False\ Postives} \quad (2)$$

$$Accuracy = \frac{True\ Positive + True\ Negative}{True\ Positives + True\ Negatives + False\ Negatives + False\ Postive} \quad (3)$$

True Positive (TP) is number of images that were correctly classified as having DR, True Negative (TN) is number of images that were correctly classified as not having DR while False Positive (FP) is number of images which weren't predicted with DR but were actually found to have DR (Type 1 Error) and False Negative (FN) is defined as the number of images which were predicted to have DR but did not actually have DR (Type 2 Error).

## VI. PROPOSED METHOD

### A. Image Preprocessing and Degradation

Various image operations were performed on the training dataset before giving it as an input for training. Resizing is the first step in this process, RGB images in the training dataset was resized to 128x128. Secondly, we rotated images to angles randomly between 0 and 360 degrees, horizontal flipping of the image, rescaling and width shift was performed. We also used Keras' preprocess_input function which is unique for each of the above-mentioned models. Figure 3 shows the preprocessed images with severity levels of DR.

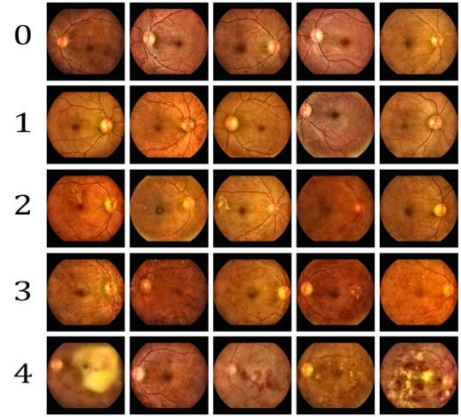

Fig. 3. Rows showing preprocessed images of each grade

To ensure that the models can perform well in a real-world environment, degradation of the training images was done. This was achieved by degrading the images belonging to the IDRiD dataset, based on multiple different combinations of Light Transmission Disturbance, Image Blurring and insertion of Retinal Artifacts which imitates the real-world disturbances as mentioned in [15]. Degradation was only applied on the IDRiD dataset increasing the total number of images from 516 to 4128.

As the images belonging to eyePACS and DDR datasets were noisy and blurred, addition of degradation on the images would lead to loss of information, whereas the images of the IDRiD dataset were of high quality so addition of degradation will ensure that they generalize in a real-world environment.

### B. Modelling

Modelling was done using CNNs, we built our models namely, Inception V3, InceptionResNetV2 and ResNet-50 which were initialized with ImageNet weights and tested on DDR and eyePACS datasets. We have used attention mechanism [16] here by putting it on top of the pretrained models.

Attention can be interpreted as a vector of importance weights which estimates how strongly a pixel/feature correlates with other and the attention vector weighted sum of feature values is taken as the approximation of the target. Attention mechanism is applied to give more importance to some of the locations of the features of the image as compared to others like MA, HM, hard and soft EX.

The models were able to classify the images according to their severity grade levels, the classification was done using the "SoftMax" classifier for multi-class classification used in the output layer of our models.

Two types of training dataset were constructed which are explained below briefly:

- IDRiD Degraded Dataset: This dataset consists of total 4128 number of images belonging to the IDRiD dataset, which were degraded with the above-mentioned techniques and were validated using 20% of images and rest were fed into the models for training.
- Balanced Combined Dataset: This dataset is a balanced version of the combination of the three datasets wherein, each label consists of 1782 images. It consists of all the 4128 images from IDRiD Degraded, 1231 images from DDR and 3551 images from eyePACS. Total 8910 images.

After training the models with the above explained datasets, they were tested on eyePACS and DDR dataset.

### C. Browser-based Application

Computer Aided Diagnosis (CAD) was built that can be used by various hospitals and clinics around the world. The browser application takes in a retinal fundus image and uses deep learning models to predict the DR grades, also it shows how sure the model is about its prediction through the probability distribution across different grades.

The browser application is also equipped with providing necessary answers as to which part of the image does model considers important for making the final decision. This is done via implementing Integrated Gradients (IG) [17]. IG Attribution Mask is overlaid with the input retinal fundus image as shown in Figure 4. This can then be used by doctors to better understand the final prediction, and also helps in understanding the mistakes the model is making while predicting. The browser application also has features to support data collection on a daily basis, after each prediction the application prompts the doctors to predict the grading, which they think is correct. This will help in Active Learning and further improve the predictive powers of the model.

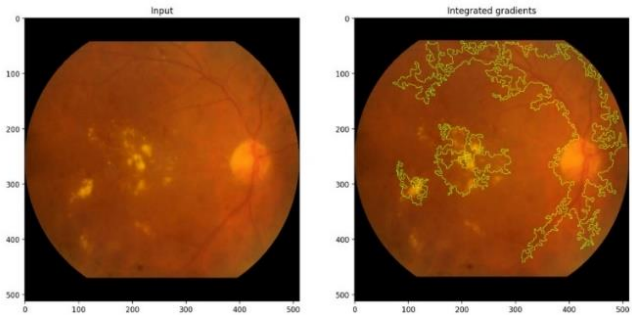

Fig. 4. Integrated Gradients Attribution Mask

The application was built using Streamlit, an easy-to-use framework build specially to support fast development of AI based applications. The main features of the application are:

1. Support for using custom build models. The application provides an admin dashboard that can be used to replace/add deep learning models very easily.
2. An easy-to-use interface for the user to predict the DR grade using the application. The end user can very easily use the deep learning models without any hassle. Ease of use was the key component taken into consideration while building the browser application.
3. Use of explainable AI in the application so as to aid Doctors and Clinicians in making the final decision on the DR grade. The application displays the probability distribution and also overlays the Integrated Gradient attribution mask on the input image. This can aid in understanding the reasons as to why the model has predicted the grade and help understand reasons as to why the final prediction was incorrect.
4. An easy way to collect data for an Institution for further enhancement of the models. We provide a module that can be added to the application. This will prompt the user (Doctor/Clinicians) to predict the correct label according to them. This can then be pooled into a server or database for storage and can be used by Active Learning algorithms to enhance the models. This is one of the key components of this browser application as model drifting is a common problem faced by deep learning models in production.

### VII. RESULTS AND DISCUSSION

In this section, we compared the Performance metrics for our models and they are tabulated in the Tables 1,2,3 and 4.

TABLE I. TRAINED ON IDRiD DEGRADED DATASET AND TESTED ON EYEPACS DATASET

| Models | Precision | Recall | AUC | Overall Accuracy |
|---|---|---|---|---|
| InceptionResNetV2 | 0.68 | 0.64 | 0.86 | 66% |
| Resnet-50 | 0.61 | 0.50 | 0.84 | 55.3% |
| InceptionV3 | 0.61 | 0.51 | 0.85 | 57% |

TABLE II. TRAINED ON IDRiD DEGRADED DATASET AND TESTED ON DDR DATASET

| Models | Precision | Recall | AUC | Overall Accuracy |
|---|---|---|---|---|
| InceptionResNetV2 | 0.55 | 0.50 | 0.80 | 52.5% |
| Resnet-50 | 0.50 | 0.50 | 0.70 | 50.01% |
| InceptionV3 | 0.40 | 0.28 | 0.76 | 36.6% |

TABLE III. TRAINED ON BALANCED COMBINED DATASET AND TESTED ON EYEPACS DATASET

| Models | Precision | Recall | AUC | Overall Accuracy |
|---|---|---|---|---|
| InceptionResNetV2 | 0.53 | 0.48 | 0.85 | 51.16% |
| Resnet-50 | 0.52 | 0.42 | 0.83 | 48.17% |
| InceptionV3 | 0.66 | 0.61 | 0.84 | 64.09% |

TABLE IV. TRAINED ON BALANCED COMBINED DATASET AND TESTED ON DDR DATASET

| Models | Precision | Recall | AUC | Overall Accuracy |
|---|---|---|---|---|
| InceptionResNetV2 | 0.31 | 0.27 | 0.76 | 30.7% |
| Resnet-50 | 0.40 | 0.22 | 0.75 | 34.02% |
| InceptionV3 | 0.31 | 0.27 | 0.76 | 30.67% |

Testing of models was done on eyePACS and DDR datasets. Using only IDRiD dataset which contains 516 images, the proposed method was able to get more than 66% accuracy and AUC over 0.85 whereas, the AUC obtained for the models trained on Balanced Combined dataset which contains total 8910 images, was in the range of 0.75 to 0.85 and best overall accuracy was 64%.

Integrated Gradients was applied to reduce the "black-box" nature of the DL algorithms and the combination of DL models and IG was used to build a browser-based application which would help in reducing the DR diagnosis time.

As the images belonging to eyePACS and DDR datasets contain noisy data, by filtering it, the model metrics can further be improved. We can extend our work by implementing Active Learning in the browser application and by using the Ensemble learning approach to build models.